\documentclass[aps,prd,twocolumn,superscriptaddress,nofootinbib]{revtex4-1}

\usepackage{amsmath}
\usepackage{amssymb}
\usepackage{hyperref}
\usepackage{color}
\usepackage{lmodern}
\usepackage{graphicx}
\usepackage{xspace}
\usepackage{xcolor}

\newcommand{\code}[1]{\texttt{#1}}

\newcommand{\mycomment}[1]{}

\newcommand{\AthenaK}{\code{AthenaK}\xspace}

\newcommand{\Msun}{\textrm{M}_\odot}
\newcommand{\ms}{\textrm{ms}}
\newcommand{\G}{\textrm{G}}

\newcommand{\km}{\mathrm{km}}

\DeclareMathOperator{\sign}{sign}

\begin{document}

\title{Magnetic Flux Emergence in Binary Neutron Star Remnants}

\author{Jacob \surname{Fields}}
\email{jmf6719@psu.edu}
\affiliation{Institute for Gravitation \& the Cosmos, The Pennsylvania State University, University Park, PA 16802}
\affiliation{Department of Physics, The Pennsylvania State University,
University Park, PA 16802}

\author{David \surname{Radice}}
\affiliation{Institute for Gravitation \& the Cosmos, The Pennsylvania State University, University Park, PA 16802}
\affiliation{Department of Physics, The Pennsylvania State University,
University Park, PA 16802}
\affiliation{Department of Astronomy \& Astrophysics, The Pennsylvania State University, University Park, PA 16802}

\author{Peter \surname{Hammond}}
\affiliation{Department of Physics and Astronomy, University of New Hampshire, Durham, NH 03824}
\affiliation{Institute for Gravitation \& the Cosmos, The Pennsylvania State University, University Park, PA 16802}
\affiliation{Department of Physics, The Pennsylvania State University,
	University Park, PA 16802}

\date{January 29, 2025}

\begin{abstract}
  Using high-resolution \AthenaK simulations of a twisted toroidal flux tube, we study the
  flux emergence of magnetic structures in the shear layer of a hot massive neutron star
  typical of a binary neutron star remnant. High-resolution simulations demonstrate that
  magnetic buoyant instabilities allow for emergence only for extremely large magnetic
  fields significantly exceeding $10^{17}~\mathrm{G}$, and more typical fields around
  $10^{16}~\mathrm{G}$ are instead dominated by hydrodynamic effects. Because merger
  remnants tend to be stable against hydrodynamic convection, our work places strong
  limitations on the mechanisms by which massive binary neutron star remnants can produce
  the magnetically-driven outflows needed to power jets.
\end{abstract}

\maketitle

\section{Introduction}
\label{sec:intro}
During a binary neutron star (BNS) merger, the shearing motion between the stellar
surfaces as they contact triggers turbulent instabilities, which help amplify the magnetic
field to magnetar strengths within the first few milliseconds after merger
\cite{Price:2006fi,PhysRevD.90.041502,Kiuchi:2015sga,Kiuchi:2017zzg,Palenzuela:2021gdo,
Aguilera-Miret:2023qih}. In the post-merger phase, this shear layer where the
Kelvin-Helmholtz instability is active settles to a region ${\sim}1~\mathrm{km}$ below the
surface \cite{Kastaun:2014fna,Hanauske:2016gia}, which is where the strongest fields are
focused \cite{Palenzuela:2021gdo, Aguilera-Miret:2023qih}. Recent simulations suggest
these strongly amplified fields can break out from the remnant star on relatively short
timescales and help power short gamma-ray bursts (sGRBs) \cite{Most:2023sft,
Combi:2023yav}. However, it is not immediately clear how this breakout occurs; BNS
remnants are typically stable against convection \cite{Kastaun:2014fna,Hanauske:2016gia,
Radice:2023zlw,Gao:2025nfj}, so this breakout cannot be purely hydrodynamical in nature.

A similar problem occurs in solar physics. Features such as coronal loops and prominences
are direct consequences of strong magnetic fields emerging from the solar surface. These
fields are generated by the solar dynamo, and they are propelled upward through the
convective zone. Their progress then stalls inside the photosphere, which is stable
against convection. A rather successful model for achieving emergence relies on magnetic
buoyancy: as magnetic flux accumulates near the base of the photosphere, the magnetized
region eventually becomes buoyant and accelerates upward
\cite{2004A&A...426.1047A}.

To illustrate how this occurs, consider a pocket of fluid with a magnetic field
$\mathbf{B}$ in pressure equilibrium with an ambient unmagnetized fluid. Therefore, the
pressure in the pocket, $P^\ast$, is $\Delta P = B^2/8\pi$ lower than the ambient
pressure, $P_0$. Assuming the pocket is either isentropic or in thermal equilibrium with
the ambient fluid, the density in the pocket will be lower than the ambient fluid and thus
experience a buoyant force.

A similar mechanism might be responsible for strong post-merger fields, which emerge from a
BNS and create magnetically-driven outflows that could potentially power relativistic
jets. A strong toroidal magnetic field develops inside the shear layer (see, e.g.,
Refs.~\cite{Aguilera-Miret:2021fre,Kiuchi:2023obe}), which could become buoyant as
it settles into hydrostatic equilibrium with the surrounding fluid. From the Newtonian
ideal magnetohydrodynamics (MHD) equations, the velocity $\mathbf{v}$ of the flux tube
evolves as
\begin{equation}
  \label{eq:newt_v_lagrange}
  \rho \frac{D\mathbf{v}}{Dt} = -\nabla\left(P + \frac{B^2}{8\pi}\right) +
                          \frac{1}{4\pi}\left(\mathbf{B}\cdot\nabla\right)\mathbf{B} +
                          \rho \mathbf{g},
\end{equation}
where $D/Dt = \partial_t + \mathbf{v}\cdot\nabla$ is the advective derivative, $P$ is the
pressure, $\rho$ is the density, $\mathbf{B}$ is the magnetic field, and $\mathbf{g}$ is
the local gravitational acceleration. Consider an ambient unmagnetized fluid with state
$\rho = \rho_0$ and $P = P_0$ in hydrostatic equilibrium. If we assume there exists a
magnetic flux tube with $\mathbf{B}\neq 0$, pressure $P^\ast$ satisfying
$P_0 = P^\ast + B^2/8\pi$ (i.e., pressure equilibrium), and density $\rho^\ast$, the
velocity is described by
\begin{equation}
  \label{eq:newt_mag_buoyancy}
  \rho^\ast \frac{D\mathbf{v}}{Dt} = \frac{1}{4\pi}\left(\mathbf{B}\cdot\nabla\right)\mathbf{B} -
                                  \Delta\rho \mathbf{g},
\end{equation}
where $\Delta\rho = \rho_0 - \rho^\ast$. For a spherically symmetric neutron star, we can
specialize to spherical coordinates and the $r$-direction. Therefore,
\begin{equation}
\begin{split}
  \label{eq:newt_vr}
  \rho^\ast \frac{D v^r}{Dt} =& \frac{1}{4\pi}\left(B^r \partial_r +
                               B^\theta\partial_\theta + B^\phi\partial_\phi\right)B^r \\
                              &+\frac{\left(B^r\right)^2 - B^2}{4\pi r} + \Delta\rho g(r),
\end{split}
\end{equation}
where we have defined $\mathbf{g} = -g(r)\mathbf{\hat{r}}$ for convenience. If we consider
$B^\phi$ to be the only non-zero component of the field (i.e., purely toroidal), then
Eq.~\ref{eq:newt_vr} simplifies to
\begin{equation}
  \label{eq:newt_vr_notwist}
  \frac{Dv^r}{Dt} = -\frac{B^2}{4\pi \rho^\ast r} + \frac{\Delta \rho}{\rho^\ast} g(r).
\end{equation}

Note the presence of the tension force $T\propto B^2/r$ on the right-hand side, which
arises because we consider a toroidal field. This differs slightly from buoyancy in solar
physics, which typically concerns relatively small sections of magnetized flux tubes. In
the solar regime, a parallel-plane approximation is appropriate (e.g.,
\cite{1993ApJ...414..357M,1996ApJ...464..999L,2001ApJ...559L..55M,2004A&A...426.1047A,
2006A&A...460..909M,2011PASJ...63..407T}), so there is no tension force at all. In the
case of a BNS remnant, however, we are interested in the large-scale circumstellar
toroidal field, and tension cannot be neglected. Recalling that $g(r) = G m(r)/r^2$ for a
spherically symmetric star, this implies that the field is only buoyant if
\begin{equation}
  \label{eq:buoyant_condition}
  B^2 < 4\pi\Delta\rho \frac{G m(r)}{r}.
\end{equation}

This appears to impose a maximum field strength on $\mathbf{B}$, but the situation is more
subtle than it appears because $\mathbf{B}$ is indirectly related to $\Delta\rho$ via
$\Delta P$. It is therefore more appropriate to think of it as a condition on the equation
of state (EOS) and permissible thermodynamic processes. Since $\Delta P = B^2/8\pi$, this
implies the condition
\begin{equation}
  2\Delta P < \Delta\rho \frac{G m(r)}{r}.
\end{equation}
Therefore, what matters is not the strength of the magnetic field, but how density changes
during the thermodynamic processes that restore pressure equilibrium; the magnetic field is
only indirectly constrained through this relationship, and it is usually a subdominant
effect. For example, if we assume a general equation of state and an isentropic flux tube,
this relation becomes (to first order)
\begin{equation}
  2\Gamma P_0 \lesssim \rho_0 \frac{G m(r)}{r},
\end{equation}
where $\Gamma = (\partial\log P/\partial\log \rho)_s$ is the (not necessarily fixed)
adiabatic index. Thus to first order, the strength of the magnetic field is not
constrained, and whether or not the magnetized fluid is buoyant instead depends on the
properties of the remnant. For a typical remnant neutron star with mass
$M \sim 2~\Msun$, radius $R\sim 10~\mathrm{km}$, and density
$\rho\sim10^{14}~\mathrm{g}/\mathrm{cm}^3$, this implies a maximum pressure of
$P_0 \sim 10^{34}~\mathrm{Ba}$, which is roughly compatible with realistic merger
conditions with $P_0 \sim 10^{33}~\mathrm{Ba}$, though less massive remnants may not be
compact enough to satisfy this requirement. However, a semi-relativistic treatment (i.e.,
strong gravity and large internal energy but $v/c \ll 1$, see
Appendix~\ref{app:gr_buoyancy}) leads to a modified buoyancy condition which is usually
less restrictive.

Note that this condition only determines buoyancy at a given moment and does not
necessarily mean a toroidal field inside a BNS remnant will emerge; similar to classic
hydrodynamic buoyancy, a stably stratified atmosphere can impede magnetic buoyancy, and
finite-size effects (e.g., drag) as the scale height shrinks can also affect buoyancy
\cite{1987SoPh..110..115S,1993ApJ...414..357M}. Nevertheless, this condition suggests that
magnetic buoyancy may be a possible mechanism for embedded magnetic fields in BNS remnants
to emerge.

Assuming the embedded magnetic field can emerge, the natural next step is to determine the
correct time scale. Though we can estimate an evolution time scale from
Eq.~\ref{eq:newt_vr_notwist}, it will be inaccurate for the length scale of interest
because of large changes in the scale height and the stratification of the remnant. A
better way to determine the time scale is to make some assumptions about the relationship
between $B^2$, $\rho^\ast$, and $P^\ast$ and integrate Eq.~\ref{eq:newt_vr_notwist} (or a
relativistic equivalent) over the neutron star profile. However, these time scales are
typically inaccurate; a time scale which neglects finite-size effects (e.g., drag forces)
will be a significant underestimate, and approximating these effects requires additional
assumptions which are not necessarily correct. Therefore, the best way to estimate the
emergence time scale for a magnetically buoyant pocket in a BNS remnant is via full
general relativistic magnetohydrodynamics (GRMHD).

In this paper, we consider an idealized model of a single flux tube embedded in the shear
layer with a magnetic field strength typical of a post-merger BNS remnant using
high-resolution (GRMHD) simulations. We begin by presenting our numerical setup in
Sec.~\ref{sec:methods}. We then present our results in Sec.~\ref{sec:results}, which
demonstrate that the flux tube only emerges for extremely strong magnetic fields well
above $10^{17}~\G$, and the dynamics of tubes with more realistic fields around
$10^{16}~\G$ are dominated by hydrodynamic effects. Next, we discuss the implications of
our work for magnetically driven outflows in realistic post-merger scenarios in
Sec.~\ref{sec:discussion}, and we finally summarize our findings and suggest avenues for
future research in Sec.~\ref{sec:conclusion}.

Though most quantities in this paper are expressed in physical units, a number of formulae
in Sec.~\ref{subsec:Bstrength} and App.~\ref{app:gr_buoyancy} instead use geometrized
units with $G=c=1$ to maintain consistency with relativistic literature. In these
scenarios, the solar mass $M_\odot$ is chosen as the relevant dimensional scale. However,
any results obtained with these formulae quoted in the text are first converted to
physical units.

\section{Methods}
\label{sec:methods}

\subsection{Initial Data}
\label{subsec:id}
We construct our model based on the results of Ref.~\cite{Radice:2023zlw}.
As shown in Figures~3, 7, and 8 of Ref.~\cite{Radice:2023zlw}, the
remnant settles into a quasi-steady state approximately $10~\mathrm{ms}$ after merger. We
perform nonlinear fits as functions of density (see Appendix~\ref{app:fits}) to the
temperature and lepton fraction profiles in order to compute a 1D finite-temperature
equilibrium slice of the HS(DD2) EOS (hereafter DD2)\cite{Typel:2009sy,HEMPEL2010210},
which we then use to generate a static neutron star via the Tolman-Oppenheimer-Volkoff
(TOV) equations \cite{Tolman2,TOV3}. Using a central density
$\rho=2.2\times10^{-3}~G^{-3}\Msun^{-2}c^6\approx
1.36\times10^{15}~\mathrm{g}/\mathrm{cm}^3$, the resulting star has a mass of
$M\approx2.43~\Msun$ and a radius $R\approx8.35~G\Msun c^{-2} \approx 12.3~\mathrm{km}$.
Because our table only covers a finite density range, we define the edge of the star as
the point dropping below $\rho=10^{-5}~G^{-3}\Msun^{-2}c^6\approx
6.18\times10^{12}~\mathrm{g}/\mathrm{cm}^3$.

Because we are interested in the dynamics of buoyant magnetic structures, we modify the
initial data, which is initially in hydrostatic equilibrium, by superimposing a toroidal
flux tube of radius $s_0$ satisfying
\begin{subequations}
\label{eq:flux_tube}
\begin{align}
  \label{eq:bphi}
  \sqrt{\gamma}B^\phi &= B_0 e^{-s^2/\sigma^2} \\
  \label{eq:bpsi}
  \sqrt{\gamma}B^\psi &= q s B_0 e^{-s^2/\sigma^2},
\end{align}
\end{subequations}
where $s^2 = r^2 + R^2 - 2Rr\sin{\psi}$ is the coordinate distance from a ring of
radius $R$ in the coordinate equatorial plane, $\psi$ is the poloidal angle of the torus,
$q$ is a twist parameter, and $\gamma$ is the determinant of the spatial metric
$\gamma_{ij}$. While this magnetic field trivially satisfies the divergence-free condition
in toroidal coordinates, $B^r$ and $B^\theta$ will both depend on $r$ and $\theta$.
Therefore, to limit numerical errors in the divergence-free condition during the
coordinate transformation, we instead compute $B^r$ and $B^\theta$ from the azimuthal
component of the vector potential, which is the same in both spherical and toroidal
coordinates (note that other contributions vanish due to axisymmetry):
\begin{equation}
\label{eq:flux_vector_pot}
  A_\phi = \begin{cases}
             -\frac{1}{2} q B_0 \sigma^2 e^{-s^2/\sigma^2} & s \leq s_0 \\
             -\frac{1}{2} q B_0 \sigma^2 e^{-s_0^2/\sigma^2} & s > s_0.
           \end{cases}.
\end{equation}

A realistic magnetic field in a merger is not completely axisymmetric, which can trigger
additional magnetic buoyant instabilities beyond the simple form of magnetic buoyancy
already discussed. Among the most important of these effects is the so-called
Parker instability (or undular instability) \cite{1966ApJ...145..811P,
1979SoPh...62...23A}. Consider a longitudinal perturbation (i.e., parallel to the
field lines) to the density or velocity to a state otherwise in equilibrium. The field
lines will bend and form a series of peaks and troughs. The fluid will attempt to slide
along the field lines, which causes the peaks to become lighter and more buoyant while the
troughs become heavier and thus fall. The Parker instability can often grow in a
stratified medium where other forms of magnetic buoyancy cannot, which makes it an
attractive prospect for enhancing buoyancy in our strongly stratified remnant profile.

Consequently, we also add a small density perturbation of the form
\begin{equation}
\label{eq:perturbation}
\rho = \rho_0\left(1-A\cos\left(k\phi\right)\right)
\end{equation}
along the length of the tube where $A$ is a specified amplitude and $k$ is an integer
determining the harmonic; because we keep the pressure fixed, this indirectly adds a
perturbation to the temperature inside the flux tube. These longitudinal perturbations add
a small amount of hydrodynamic buoyancy to the system, causing the field lines to bend and
possibly seeding the Parker instability.

After superimposing the magnetic flux tube and applying the fluid perturbation, we
enforce mechanical equilibrium with the TOV solution by setting the pressure inside the
tube to $P^\ast = P_0 - P_B$, where $P_0$ is the ambient fluid pressure and $P_B$ is
magnetic pressure ($P_B = b^2/8\pi$ in CGS units, where $b^\mu$ is the magnetic field in
the fluid's rest frame). The density, which has already been perturbed, is then adjusted
again to keep the (perturbed) temperature fixed. The composition of the fluid is also held
constant during this process. Therefore, the fluid is in pressure equilibrium with the TOV
solution and thermal equilibrium with the perturbed state, but it deviates slightly from
thermal equilibrium with the TOV (leading to hydrodynamic buoyancy), and it cannot
generally achieve hydrostatic equilibrium due to the magnetic field. In a true post-merger
scenario, the thermal diffusion time scale is very long compared to the dynamical time
scale, so calculating the density difference assuming an adiabatic process rather than an
isothermal one may produce a more realistic flux tube. However, one can show that
enforcing thermal equilibrium results in a larger density difference than an isentropic
system, independent of the EOS, because
$(\partial \rho/\partial P)_T > (\partial \rho/\partial P)_s$ is equivalent to requiring
the heat capacities to satisfy $C_P/C_V > 1$. Therefore, these tests tend to overestimate
how buoyant a flux tube is and provide something of a best-case scenario for emergence.

\begin{table}
\begin{ruledtabular}
  \caption{\label{tab:ics} A summary of the simulations performed, including the initial
    maximum field strength, the radius $s_0$ of the flux tube, the estimated emergence
    time, and the estimated velocity at emergence.}
    \begin{tabular}{l|cccc|cc}
    Name & $\sqrt{b^2_\mathrm{max}}$ (G) & $A$ & $s_0$ (km) & k &
        $t_\mathrm{em}$ (ms) & $v_\mathrm{em} (c)$ \\
    \hline
    H7e17\_5         & $7.540\times10^{17}$ & $0.05$ & $0.369$ & 18 & 0.216 & 0.038 \\
    H7e17\_5\_B      & $7.540\times10^{17}$ & $0.05$ & $1.48$  & 18 & 0.201 & 0.013 \\
    H7e17\_5\_k9     & $7.540\times10^{17}$ & $0.05$ & $0.369$ & 9  & 0.194 & 0.048 \\
    H7e17\_0         & $7.540\times10^{17}$ & $0$    & $0.369$ & N/A & 0.224 & 0.034 \\
    H1e17\_5         & $1.257\times10^{17}$ & $0.05$ & $0.369$ & 18 & N/A  & N/A  \\
    H1e17\_5\_k9     & $1.257\times10^{17}$ & $0.05$ & $0.369$ & 9  & N/A  & N/A  \\
    H1e17\_0         & $1.257\times10^{17}$ & $0$    & $0.369$ & N/A & N/A  & N/A  \\
    H1e16\_5         & $1.257\times10^{16}$ & $0.05$ & $0.369$ & 18 & N/A  & N/A  \\
    H6e15\_5         & $6.283\times10^{15}$ & $0.05$ & $0.369$ & 18 & N/A  & N/A  \\
    H0\_5            & $0$                  & $0.05$ & $0.369$ & 18 & N/A  & N/A  \\
    \end{tabular}
\end{ruledtabular}
\end{table}

For most of our runs, we consider a flux tube of inner radius
$s_0 = 0.25~G\Msun c^{-2}\approx369~\mathrm{m}$ and an outer radius $R$ matching the
position of the peak temperature ($r=7.114~\Msun$), which we take as a proxy for the
position of the shear layer. We additionally set the width parameter $\sigma = 0.3$ and
add a small twist with $q=1~G^{-1}\Msun^{-1}c^2\approx0.677~\km^{-1}$ to increase rigidity
of the flux tube. The perturbation wavenumber is set to $k=18$.

Across our runs we vary the magnetic field strength and the perturbation amplitude. Based
on the relativistic generalization of Eq.~\ref{eq:buoyant_condition}
(Eq.~\ref{eq:gr_inequality}), only very weak fields ($B\lesssim 4\times10^{12}~\G$) are
completely incapable of buoyancy. Due to the finite width of the flux tube and
discretization effects, the necessary field strength for buoyancy can deviate somewhat
from this ideal condition, but it serves as a useful guideline nonetheless. We test fields
at $7.540\times10^{17}~\G$, $1.257\times10^{17}~\G$, $1.257\times10^{16}~\G$, and
$6.283\times10^{15}~\G$. For every field strength, we perform a run with a perturbation
amplitude $A=0.05$. We perform an additional test with $A=0$ for some cases to help
investigate the role that the Parker instability plays in the perturbed runs. As a
control, we also perform a run with no magnetic field but with an $A=0.05$ density
perturbation; this will help disentangle the effects of ordinary hydrodynamic buoyancy
from those caused by the Parker instability. To help track the evolution of the tube for
this case, a passive tracer scalar $X$ is added with $X=1$ inside the tube and $X=0$
everywhere else.

Each run is recorded in Table~\ref{tab:ics}, with the names following the format
``Px\_A''. ``P'' is a letter indicating the stellar profile, ``x'' indicates the
magnetic field with the leading digit and order of magnitude, and ``A'' is the strength
of the amplitude in percent. Thus the run with a $7.540\times10^{17}~\G$ field embedded in
the hot remnant profile with an $A=0.05$ perturbation is labeled ``H7e17\_5``, while
``H0\_5'' corresponds to the base case with a density perturbation but no magnetic field.

Since the shear layer is ${\sim}3~\km$ wide in a realistic BNS merger
\cite{Hanauske:2016gia}, a flux tube of $s_0 = 0.25~G\Msun/c^2$ may be much smaller in size
than the toroidal field structure which develops in a realistic merger simulation. Because
we may approximate the field as a superposition of weakly twisted flux tubes, tests with
small flux tubes should be qualitatively correct when determining if a particular flux
tube can emerge. Nevertheless, we do include one additional setup with the H7e17\_5
configuration with a larger flux tube as well. For this run, we set
$s_0 = 1.0~G\Msun/c^2\approx1.48~\km$ and reduce the twist parameter to
$q=0.25~G^{-1}\Msun^{-1}c^2\approx0.169~\km^{-1}$; since the twist scales linearly with
the tube radius $s$, this is necessary to ensure the magnetic field remains primarily
toroidal. We label this run H7e17\_5\_B to distinguish it from the case with a small flux
tube.

We also consider two additional runs similar to H7e17\_5 and H7e17\_1 but with $k=9$,
which we label as H7e17\_5\_k9 and H1e17\_5\_k9, respectively, to better understand any
possible role of the Parker instability, which is wavenumber-dependent. In a
plane-parallel approximation in Newtonian MHD, the maximum growth rate occurs at
$\lambda \sim 10 H_P$, where $\lambda$ is the wavelength and $H_P$ is the pressure scale
height \cite{1988PASJ...40..147H}. This corresponds to $k\approx8$ for our configuration,
so we choose $k=9$ as it is the smallest perturbation that fits a full wavelength inside
our domain. We acknowledge that our system is neither plane-parallel nor Newtonian, but we
expect that the leading-order corrections to the growth rate from GR and a toroidal
geometry should be similar across all wavenumbers. Note also that $k\approx8$ is only the
critical wavenumber near the initial location of the flux tube. As the flux tube rises,
the scale height will rapidly decrease, so the critical wavenumber will increase. The
``correct'' perturbation for maximum growth, then, depends on the timescale for the Parker
instability during the linear phase. If this growth rate is much shorter than the
emergence timescale, the linear phase will be dominated by the initial conditions. At the
other extreme, where the timescale is much longer than the emergence timescale, the
Parker instability will not play a significant role in the dynamics, so the perturbation
will be irrelevant.


We note that our strongest fields are unrealistically large for a BNS post-merger;
high-resolution BNS simulations starting with realistic initial field strengths do not
typically produce fields with peak strengths much in excess of $10^{17}~\G$, and average
strengths are typically ${\sim}10^{16}~\G$ \cite{Kiuchi:2015sga,Kiuchi:2017zzg,
Mosta:2020hlh,Palenzuela:2021gdo,Aguilera-Miret:2021fre,Kiuchi:2023obe}. Though no
full-scale merger simulation has sufficient resolution to capture all scales of magnetic
turbulence, local simulations confirm these findings \cite{Obergaulinger:2010gf,
Zrake:2013mra}.

Note also that the remnant in Ref.~\cite{Radice:2023zlw} is differentially rotating, but
we have neglected these effects by using a TOV solution. Though rotation can modify
buoyancy and convective instabilities, it tends to inhibit these mechanisms
\cite{1978RSPTA.289..459A,Spruit:1999cc,Gao:2025nfj}. In conjunction with our assumption
of a flux tube in thermal (rather than isentropic) equilibrium, our results will tend
toward overestimating the buoyancy of the system.



\subsection{Numerical Methods}
\label{subsec:num_methods}
We evolve the star with the GRMHD code \AthenaK \cite{Stone.6.24,Zhu:2024utz,
Fields:2024pob}, which uses a second-order finite-volume method to evolve the fluid and
upwind constrained transport to ensure that $\nabla\cdot\mathbf{B}=0$. For these tests, we
choose WENOZ as our reconstruction scheme \cite{2008JCoPh.227.3191B} and enable a
first-order flux correction (FOFC) method \cite{Lemaster:2008gh}. The FOFC scheme greatly
improves the stability of the evolution near the surface of the neutron star, particularly
when the magnetization is high. Because we assume that the remnant has achieved steady
state, we fix the spacetime to the initial TOV solution. Because the energy density inside
the flux tube is reduced, there will be a gravitational mass deficit that should affect
the metric. However, even for the H7e17\_5\_B case, this deficit is $\lesssim 2.2\%$ (and
$\lesssim 0.2\%$ for H7e17\_5), so this effect is a small perturbation that can safely be
ignored.

Our computational domain is a wedge in the equatorial plane covering
$r \in [6~G\Msun c^{-2}, 24~G\Msun c^{-2}]\approx[8.86~\km,35.44~\km]$,
$\theta \in [\pi/2 - \pi/9, \pi/2 + \pi/9]$, and $\phi \in [-\pi/9, \pi/9]$, which amounts
to full two oscillations for the $k=18$ perturbation described in Sec.~\ref{subsec:id}. We
use a base domain of $600\times180\times180$ cells with a single refinement region over
$r\in[6~G\Msun c^{-2},12~G\Msun c^{-2}]\approx[8.86~\km,17.72~\km]$ for an effective
radial resolution of $\Delta r = 0.015~G\Msun c^{-2} \approx 22~\mathrm{m}$ inside the
star. The angular resolution is sufficient that
$r\Delta\theta \approx r\sin{\theta}\Delta\phi \approx \Delta r$ over the region of
interest from approximately $7~G\Msun c^{-2}\approx10~\km$ to
$9~G\Msun c^{-2}\approx13~\km$.

The upper boundary of $r$ is a ``diode'' outflow boundary (the radial momentum must be
nonnegative), which limits spurious fluxes generated at the level of the artificial
atmosphere from inflowing back into the computational domain. The inner boundary of $r$ is
designed to approximate a ``stress-free'' boundary. We first fix the rest-mass density
$\rho$ and pressure $P$ to the initial TOV solution. The radial velocity $W v^r$, where
$W = (1-v^2)^{-1/2}$ is the Lorentz factor and $v^i$ is the three-velocity, is then
determined by setting $F_\mathrm{in} = -F_\mathrm{out}$ for the mass flux
$F=\alpha\sqrt{\gamma}\rho W v^r$, where $\gamma$ is the determinant of the spatial metric
and $\alpha$ is the lapse, and solving for $(W v^r)_\mathrm{in}$. This behaves nearly like
a typical reflection boundary, but it corrects for $\alpha\sqrt{\gamma}\rho$ being
slightly larger on the inside of the star. The magnetic field and $W v^\phi$ and
$W v^\theta$ are also reflected. Note, however, that we invert the parity for all
components rather than only the radial component. In the event that the flux tube comes
too close to the inner boundary, we found that a simple reflection that does not invert
the parity of the tangential components of the magnetic field and velocity can cause small
numerical errors to accumulate in the velocity and generate an unphysical rotation in the
star. Through experimentation, we found that also flipping the tangential components
damps these errors out and greatly improves stability. In practice the flux tube remains
far from the radial boundary in all the runs presented here, so this condition has
little effect on the evolution.

We exploit the spherical symmetry of the TOV solution and apply periodic boundaries in
both the $\theta$ and $\phi$ directions. Though this means there are multiple copies of
the flux tube along the polar direction, this is only an issue if the flux tube expands
enough to interact with the polar boundary. So long as the flux tube is sufficiently
resolved, this only happens some time after the flux tube emerges from the star; we end
our runs before this occurs.

This work represents the first use of 3D EOS tables in \AthenaK, which are implemented
following the same prescription as in \code{GR-Athena++} \cite{Cook:2023bag}. To
approximate their effects on the remnant, we modify the full DD2 EOS table with the
contributions of trapped neutrinos following the prescription in
Ref.~\cite{Perego:2019adq}. Therefore, the evolved species fraction is now the (electron)
lepton fraction, $Y_l = Y_e + (n_{\nu_e} - n_{\nu_{\overline{e}}})/n_b$, which is advected
along with the fluid flow. As $Y_e$ becomes a function of $Y_l$, $n_b$, and the
temperature $T$, this prescription allows us to approximate the effects of trapped,
fast-equilibriating neutrinos without needing to evolve them directly.

We evolve each simulation either until shortly after the flux tube emerges or until it
becomes clear that the flux tube will not emerge; the exact length depends on the initial
conditions and the observed dynamics.

\section{Results}
\label{sec:results}
\begin{figure}
  \centering
  \includegraphics[width=\columnwidth]{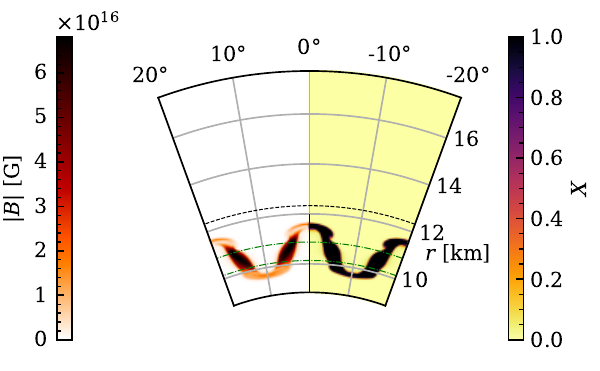}
  \caption{\label{fig:hbuoy} A slice of the $r$-$\phi$ plane at $t=0.39~\ms$ comparing
    $|B|$ from H1e16\_5 (left) to the tracer scalar $X$ from H0\_5
    (right). The dashed black line marks the surface of the star, and the dot-dashed green
    lines mark the boundaries of the initial position of the flux tube. The nearly
    identical evolution indicates that the dynamics of H1e16\_5 are driven primarily by
    hydrodynamic buoyancy from the original perturbation rather than magnetic buoyancy.}
\end{figure}

We begin by discussing the non-magnetized base case, H0\_5. As shown in
Fig.~\ref{fig:hbuoy}, the density perturbation causes lighter parts of the tracer region
to rise while the overdense regions sink, in agreement with standard hydrodynamic
buoyancy. However, because the remnant is stable against convection, the perturbations
slow down and eventually settle into layers of the remnant with similar densities by
$t=0.39~\ms$.

\subsection{Effects of magnetic field strength}
\label{subsec:Bstrength}
The H6e15\_5 and H1e16\_5 cases, which feature the weakest magnetic fields, are
qualitatively similar to the H0\_5 case. From comparing the magnetic field of H1e16\_5 to
the tracer of H0\_5 in Fig.~\ref{fig:hbuoy}, it is clear that the fluid within the flux
tube rises and falls exactly the same way as predicted by hydrodynamic buoyancy alone.
This is not surprising, as the magnetization $\sigma_b = b^2/4\pi\rho c^2$ is initially
${\sim}10^{-5} \ll \Delta\rho/\rho$. The magnetic field does appear to become more
concentrated in between the peaks and troughs, increasing by a factor of ${\sim}5$ in this
brief time.

To determine if this is dynamically important, we examine the energetics of
the system. We start with the stress-energy tensor for ideal MHD,
\begin{equation}
  T^{\mu\nu} = (\rho h + b^2)u^\mu u^\nu +
    \left(P + \frac{b^2}{2}\right) g^{\mu\nu} - b^\mu b^\nu,
\end{equation}
where $\rho$ is the rest-mass density, $h$ is the total specific enthalpy, $b^\mu$ is the
magnetic field in the fluid frame, $P$ is the fluid pressure, and $u^\mu$ is the
four-velocity. Also note that we have switched from the Gaussian convention for
electromagnetism to Lorentz-Heaviside to better align with the form of the GRMHD equations
as used in evolution codes. In the Valencia formalism used in \AthenaK, we evolve the
energy as $\tau = \alpha^2 T^{00} - \rho W$, where $\alpha$ is the lapse and $W$ is the
Lorentz factor, which is equivalent to
\begin{equation}
  \tau = \rho W\left(hW - 1\right) + B^2 - P - \frac{b^2}{2}.
\end{equation}
It is easy to verify in the Newtonian limit that this reduces to the usual Newtonian
expression for the total non-gravitational energy in MHD. This expression has
contributions from the internal, kinetic, and magnetic energy of the fluid. Similar to
Ref.~\cite{Cook:2025zzy}, we express the kinetic energy $E_K$ and magnetic energy $E_B$ as
\begin{align}
  E_K &= \int \rho W \left(W - 1\right) \mathrm{d}V, \\
  E_B &= \int \left(B^2 - \frac{b^2}{2}\right) \mathrm{d}V,
\end{align}
where $dV$ includes the appropriate volume element. Since $\tau$ does not include the
gravitational energy of the system, it is not conserved. However,
$-\alpha (T^0_0 + \rho u^0)$ does include these contributions and is therefore conserved
in stationary spacetimes. It is easy to verify that
$\alpha T^0_0 + \alpha\rho u^0 + \tau \approx -\rho \Phi$ in the Newtonian limit, where
$\Phi$ is the gravitational potential. We therefore define
\begin{equation}
  E_\mathrm{tot} = -\int \left(T^0_0 + \rho u^0\right) \alpha \mathrm{d}V,
\end{equation}
which we use to normalize the energy of our system. We caution that $E_\mathrm{tot}$ is
not precisely invariant due to boundary effects and truncation error from evolving $\tau$
instead of $\alpha (T^0_0 + \rho u^0)$, leading to a maximum deviation of ${\sim}0.5\%$ in
the total energy over the course of the evolution. However, we have confirmed that our
results are the same if we normalize by $E(t=0)$ or $E(t)$, indicating that these
conservation violations are insignificant for our analysis.

\begin{figure}
  \includegraphics[width=\columnwidth]{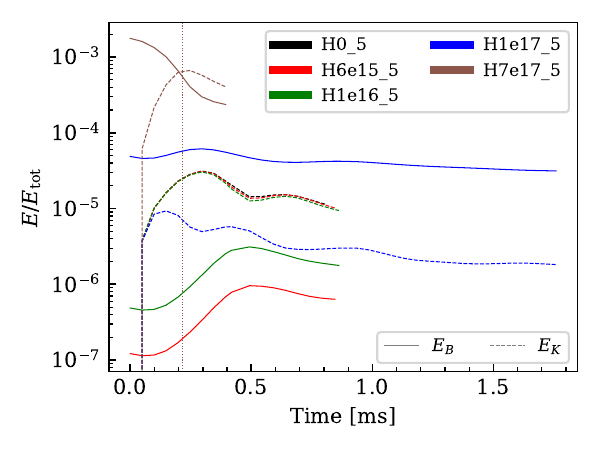}
  \caption{\label{fig:energetics} The magnetic energy $E_B$ (solid lines) and kinetic
  energy $E_K$ (dashed lines) as a function of time normalized by the total energy for
  all the runs with perturbations. The dotted vertical line marks the emergence time for
  H7e17\_5.}
\end{figure}

In Fig~\ref{fig:energetics}, we see that $E_B \ll E_K$ for both H6e15\_5 and H1e16\_5, and
$E_K$ deviates only slightly from H0\_5 after ${\sim}0.3~\ms$. The magnetic energy grows
steadily until ${\sim}0.5~\ms$, then begins to decay with $E_K$. The simultaneous decay of
both these quantities suggests that both systems are stable and are converging toward a
new equilibrium.

\begin{figure}
  \centering
  \includegraphics[width=\columnwidth]{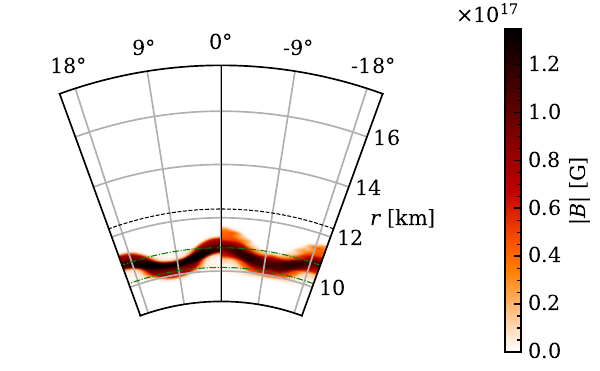}
  \caption{\label{fig:period} Snapshots of $\tilde{B}^\phi$ in the $r$-$\phi$ plane from
    H1e17\_5 at $t=0.39~\ms$ (left) and $t=1.31~\ms$ (right). The location of the flux
    tube is roughly the same in both, but some expansion and diffusion is apparent at the
    later time. The dashed black line indicates the location of the stellar surface from
    the TOV solution, and the dot-dashed green lines mark the boundaries of the flux tube
    at $t=0$.}
\end{figure}

The H1e17\_5 almost immediately shows deviations from the hydrodynamic case. Due to the
increased magnetic tension relative to the perturbation, the flux tube does not bend as
readily, and its peaks rise only a short distance before stalling and oscillating in
place. Fig.~\ref{fig:period} shows the flux tube at both $t=0.39~\ms$ and $t=1.31~\ms$ to
demonstrate this; though there is some expansion and diffusion apparent in the flux tube
at later times, the location of the core is nearly unchanged.

Fig.~\ref{fig:energetics} corroborates this picture; for H1e17\_5, $E_K$ peaks at a lower
energy and begins to decay sooner, indicating that this case is also stable. The magnetic
energy decays as well, albeit more slowly than the kinetic energy. If we consider a volume
$V = A \ell$, where $A$ is some changing cross-sectional area and $\ell$ is an
approximately fixed length, flux conservation implies that
$2E_B \approx B^2 V \propto A^{-1}$, so we attribute this slow decay in magnetic energy to
the slow diffusion or expansion of the flux tube as demonstrated in Fig.~\ref{fig:period}.

\begin{figure*}
\centering
  \includegraphics[width=\textwidth]{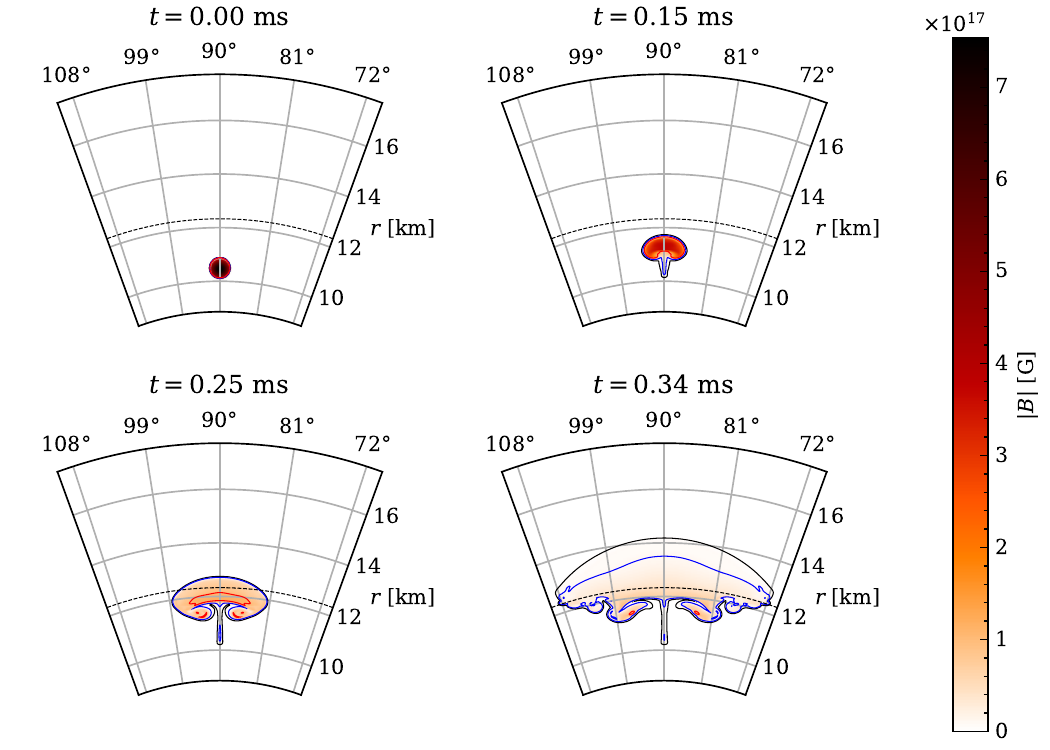}
  \caption{\label{fig:evolution} The magnitude of the magnetic field $|B|$ in the
  $r$-$\theta$ plane from H7e17\_1\_5 at $t=0~\ms$ (top-left), $t=0.15~\ms$ (top-right),
  $t=0.25~\ms$ (bottom-left) and $t=0.34~\ms$ (bottom-right). The contours in black,
  blue, and red denote $10^{15}~\G$, $10^{16}~\G$, and $10^{17}~\G$, respectively. As the
  flux tube rises, it expands into the umbrella shape characteristic of a stratified
  environment with a decreasing scale height. The dashed black line marks the location of
  the stellar surface from the TOV solution.}
\end{figure*}

\begin{figure*}
  \centering
    \includegraphics[width=0.45\textwidth]{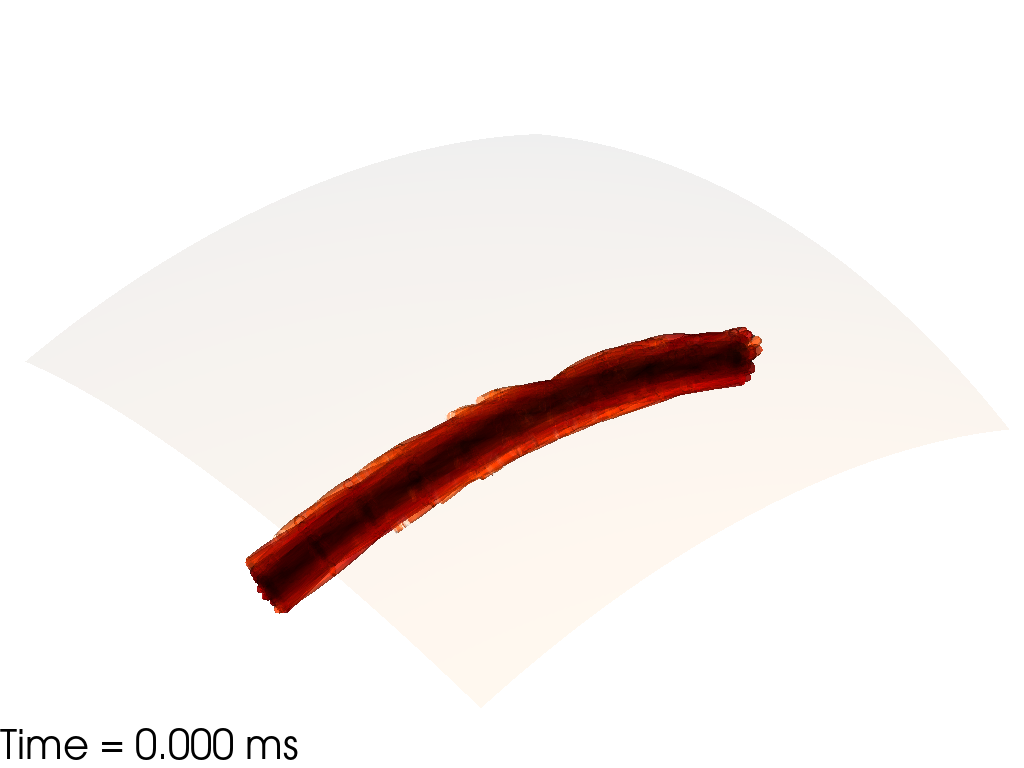}
    \includegraphics[width=0.45\textwidth]{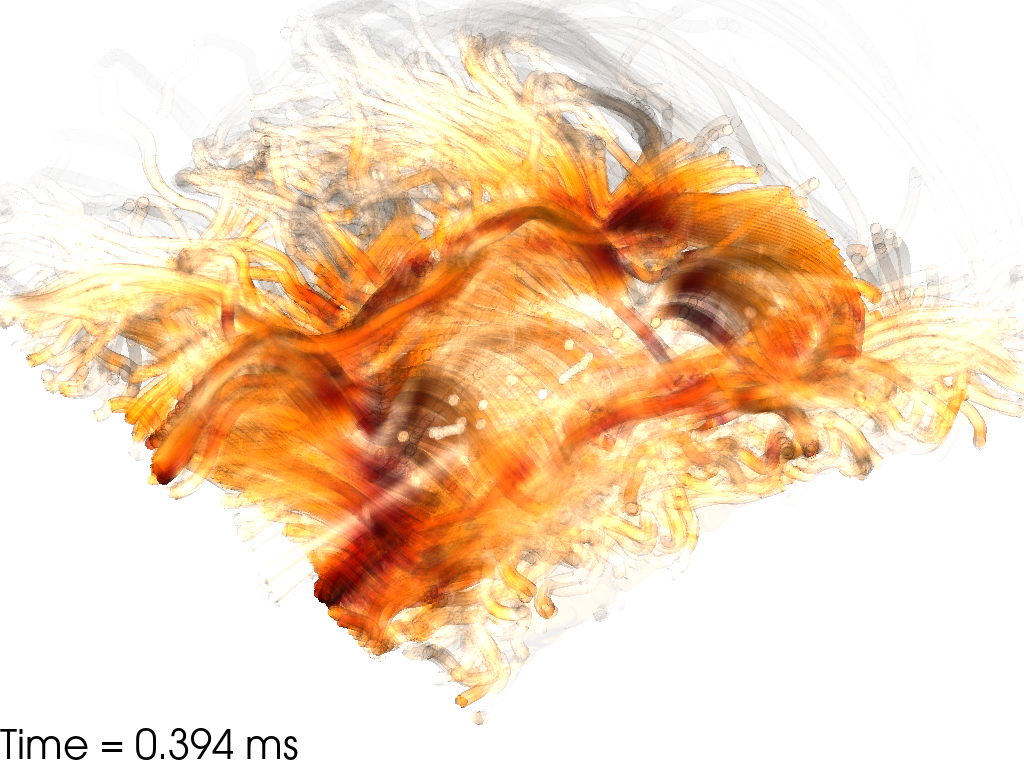}
    \includegraphics[width=0.45\textwidth]{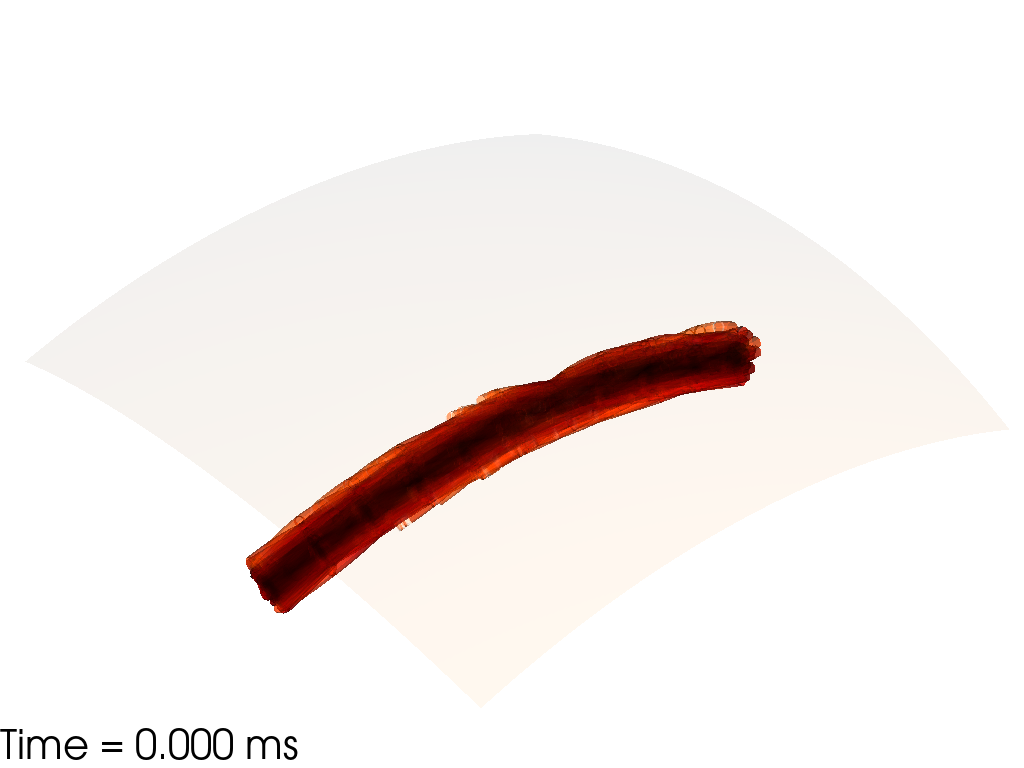}
    \includegraphics[width=0.45\textwidth]{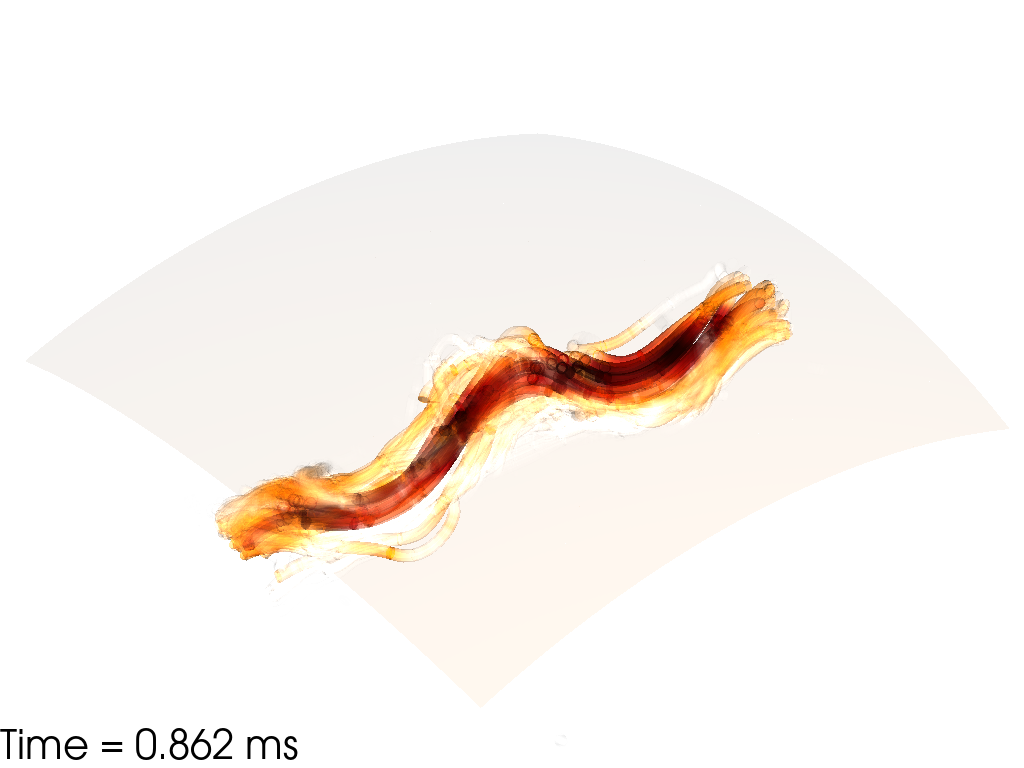}
    \caption{\label{fig:streamlines} Streamlines of the magnetic field for H7e17\_5 (top)
      and H1e17\_5 (bottom) at the initial time (left) and a late time (right). The
      surface contour marks
      $\rho=10^{-5}~\Msun^{-2}\approx6.2\times10^{12} \mathrm{g}/\mathrm{cm}^3$, which is
      the approximate location of the surface. Note that the color scale is normalized to
      the maximum field strength in each run. H7e17\_5 expands rapidly as it emerges,
      while H1e17\_5 remains trapped beneath the surface and largely intact.}
\end{figure*}

\begin{figure}
  \centering
  \includegraphics[width=\columnwidth]{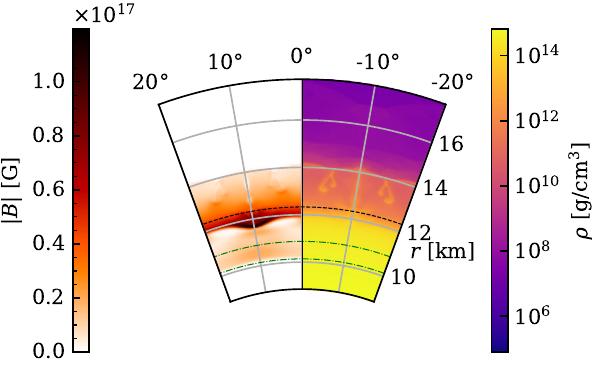}
  \caption{\label{fig:combined} A comparison of $|B|$ (left) and $\rho$ (right)
  in the $r$-$\phi$ plane from H7e17\_5 at $t=0.34~\ms$. Much of the flux tube
  remains near the surface, but the wave of material pushed out by it is also visible.
  The dashed black line marks the location of the stellar surface from the TOV solution,
  and the dot-dashed green lines mark the boundaries of the flux tube at $t=0$.}
\end{figure}

For H7e17\_5, which has the strongest magnetic field, the flux tube successfully emerges.
This emergence process is shown in Figs.~\ref{fig:evolution} and \ref{fig:streamlines}.
The flux tube stretches out into a characteristic umbrella shape due to the decreasing
scale height as it rises, similar to what occurs in solar studies (e.g., Fig.~4 of
Ref.~\cite{2011ApJ...735..126T}). As it reaches the surface, the flux tube rapidly expands
and pushes out a wave of material in front of it (see Fig.~\ref{fig:combined}).

To measure the emergence time and properties of the entrained material, we integrate the
mass flux $\dot{M}$ across surfaces at $r=8.35~G\Msun c^{-2}\approx12.33~\km$,
$r=8.4~G\Msun c^{-2}\approx12.40~\km$, $r=8.5~G\Msun c^{-2}\approx12.55~\km$,
$r=8.6~G\Msun c^{-2}\approx12.70~\km$, and $r=8.7~G\Msun c^{-2}\approx12.85~\km$. We
define the emergence time as the peak mass flux $\dot{M}$ across the surface at
$r=8.35~G\Msun c^{-2}$, which is the approximate location of the remnant's surface. By
looking for where $\dot{M}$ peaks in the other integration surfaces, we can measure the
approximate speed of the flux tube at emergence, which we do by fitting a quadratic
polynomial for $r(t)$ to these peak times and taking its time derivative.

\begin{figure}
  \centering
  \includegraphics[width=\columnwidth]{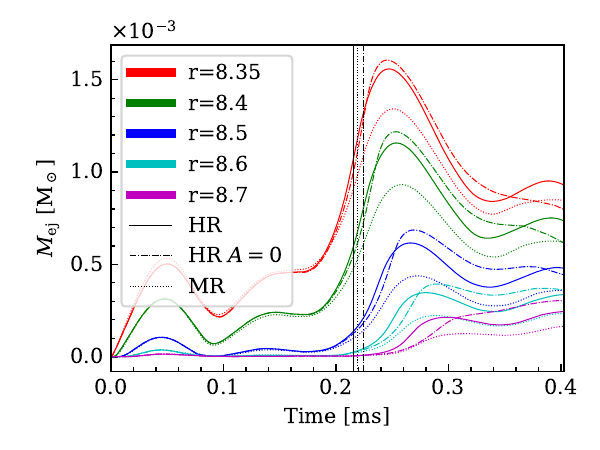}
  \caption{\label{fig:ejecta} Total integrated mass flux for H7e17\_5 at high (solid
    lines) and medium resolution (dotted lines). The integrated flux is also shown for
    H7e17\_0 at high resolution (dash-dotted line). The low-resolution data for
    H7e17\_5 is not shown because it does not emerge. Black vertical lines indicate
    emergence time as estimated by the peak instantaneous flux across the first
    integration surface at $r=8.35~\Msun$.}
\end{figure}

Fig.~\ref{fig:ejecta} shows the time-integrated mass across each integration surface along
with the estimated emergence time. The flux tube emerges at
$t_\mathrm{em}\approx0.216~\ms$ for H7e17\_5 with a speed of $v_\mathrm{em}\approx0.038c$.
The total entrained mass peaks around $M\approx1.5\times10^{-3}~\Msun$ shortly after
emergence, but it rapidly declines as material falls back onto the remnant. The mass
across each integration surface also rapidly decreases with distance, suggesting that
little to no mass is unbound\footnote{A better way to estimate unbound ejecta is using the
geodesic and/or Bernoulli criteria, but this must be done farther away from the remnant to
be accurate. Because our simulations only span $[-\pi/9,\pi/9]$ in the $\theta$ direction
with periodic boundary conditions, the emerging fluid begins to interact with itself not
long after passing the surface at $r=8.7~G\Msun c^{-2}$. Therefore, the dynamics of the
fluid will not be reliable at a radius where a proper ejecta calculation would be
meaningful.}.

Unsurprisingly, Fig.~\ref{fig:energetics} demonstrates that H7e17\_5 has larger $E_B$ and
$E_K$ than in the weak-field runs. $E_K$ follows the same qualitative trend as the other
runs, with a rapid rise followed by a peak and slow decay in time (possibly with small
oscillations). Unlike the other runs, however, $E_B$ immediately decays, while it remains
nearly constant in H1e17\_5 and grows for some time in both H1e16\_5 and H6e15\_5. There
is also a crossover between $E_B$ and $E_K$ at the same approximate time emergence occurs,
though, as will be discussed later, this may simply be coincidence.

\subsection{Effects of density perturbation}
\begin{figure}
  \centering
  \includegraphics[width=\columnwidth]{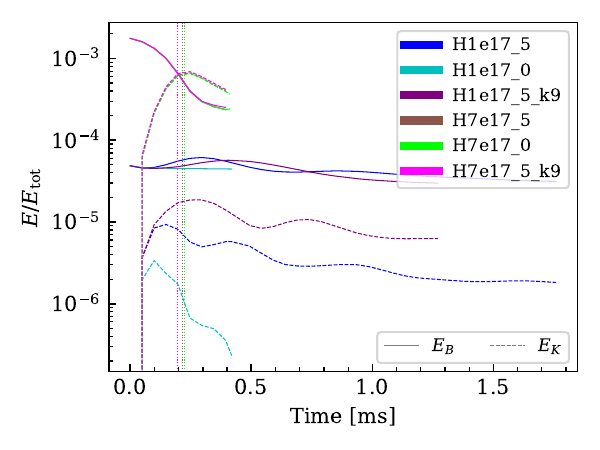}
  \caption{\label{fig:energetics_A0} Like Fig.~\ref{fig:energetics}, but comparing the
    cases with different perturbations ($A=0$, $k=9$, and $k=18$). The dotted vertical
    lines mark the emergence times for H7e17\_0, H7e17\_5, and H7e17\_5\_k9.}
\end{figure}

We now consider the effects of the perturbation by analyzing the non-perturbed cases and
the $k=9$ cases, beginning with H1e17\_0. Since H1e17\_5, which has a density
perturbation, does not emerge, it is not surprising that H1e17\_0 does not, either, and
instead rises only for a short time before stalling. As shown in
Fig.~\ref{fig:energetics_A0}, $E_K$ peaks at ${\sim}1/3$ its value in H1e17\_5, and it
rapidly decays by nearly an order of magnitude in ${\sim}0.3~\ms$. Additionally, though
the evolution of $E_B$ in H1e17\_0 is initially similar to H1e17\_5, it does not
experience the same slow rise, but instead slowly decays over time.

Comparing the H7e17\_0 case to H7e17\_5, we find that the perturbation has very little
effect on the dynamics of the system. This is consistent with our initial conditions, as
the change in density from our equilibrium conditions on the flux tube
($\Delta\rho/\rho_0\approx0.6$) is much larger than the change in density from the
perturbation ($A=0.05$). According to Fig.~\ref{fig:energetics_A0}, the presence of a
perturbation does not affect the energetics of the system in any significant way. However,
there is a small effect on both the emergence time and the mass pushed out by the emerging
flux tube. Fig.~\ref{fig:ejecta} shows that emergence in H7e17\_0 is slightly delayed to
$t_\mathrm{em}\approx0.224$ and slightly slower at $v_\mathrm{em}\approx0.034c$ compared
to H7e17\_5, though the entrained mass is slightly larger. Nevertheless, the mass again
rapidly decays with distance, suggesting that the amount of unbound ejecta is small.

H1e17\_5\_k9 demonstrates sustained growth in the kinetic energy compared to H1e17\_5 due
and therefore rises further, which at first glance may suggest that $k=9$ may be closer to
the critical wavenumber than $k=18$. However, the flux tube still fails to emerge, and the
kinetic energy never exceeds what is observed in Fig.~\ref{fig:energetics} for H0\_0\_5.
At early times, the growth rate is nearly identical for both perturbation modes, so it
seems that the differences in evolution are not necessarily because $k=9$ initially
excites the Parker instability but cannot overcome the stratification of the system near
the surface; rather, it seems that the dynamics of both $k=9$ and $k=18$ are dominated by
the hydrodynamic perturbation, and $k=9$ suppresses it less due to reduced magnetic
tension.

\begin{figure}
  \centering
  \includegraphics[width=\columnwidth]{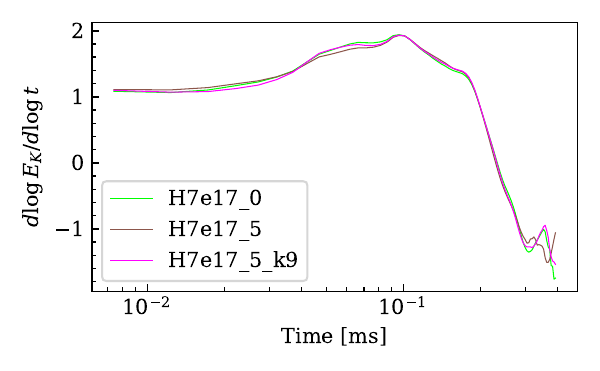}
  \caption{\label{fig:energetics_K} The growth rate $d\log E_K/d\log t$ as a function
    of time for H7e17\_0, H7e17\_5, and H7e17\_5\_k9.}
\end{figure}

On the other hand, the H7e17\_5\_k9 case emerges at $t_\mathrm{em}\approx0.194~\ms$ with
a velocity of $v_\mathrm{em}\approx0.048c$, somewhat earlier than H7e17\_5, and there
is evidence that the kinetic energy is marginally larger. When comparing the kinetic
energy evolution at early times for perturbed cases to H7e17\_0, we find a maximum
relative difference of ${\sim}9\%$ for H7e17\_5\_k9 compared to ${\sim}2\%$ for H7e17\_5.
This might suggest the $k=9$ perturbation grows faster, but we note that this difference
does not demonstrate clear monotonic growth, so we claim with any confidence that this is
more than a numerical effect. Additionally, assuming $E_K \propto t^p$ for some growth
scale $p$, we do not see any meaningful difference in $p$ between the different
perturbations. Fig.~\ref{fig:energetics_K} indicates that $d\log E_K/d\log t$ is
effectively identical for all three cases, even at early times. Therefore, the growth
timescale for the Parker instability must be much longer than the emergence timescale.
This is not truly surprising, as the Parker instability does not usually dominate over
other buoyant instabilities until it becomes nonlinear (see, e.g.,
Refs.~\cite{1993ApJ...414..357M,2005PASJ...57..995N}).

\subsection{Effects of flux tube size}
\label{subsec:size}
\begin{figure}
  \centering
  \includegraphics[width=\columnwidth]{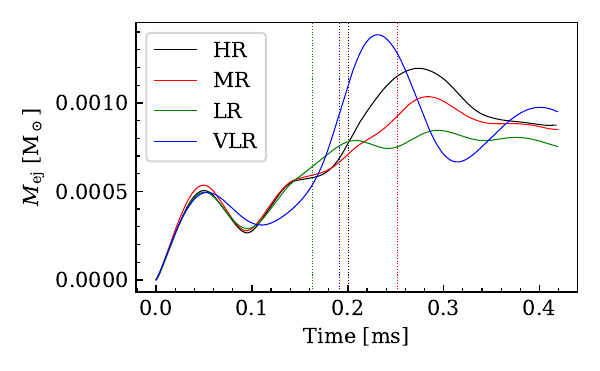}
  \includegraphics[width=\columnwidth]{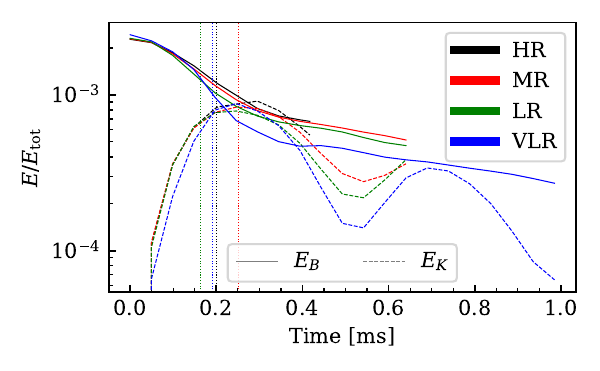}
  \caption{\label{fig:ejecta_big} (Top) The integrated mass flux across the surface of
    the star for the resolution study with H7e17\_5\_B. (Bottom) Like
    Fig.~\ref{fig:energetics}, but comparing the different resolutions of H7e17\_5\_B.
    Dotted vertical lines in both plots mark the measured emergence time using the peak of
    $\dot{M}$.}
\end{figure}

Thanks to its extended profile, the H7e17\_5\_B run emerges at ${\sim}0.201~\ms$ according
to the mass flux criterion, slightly earlier than H7e17\_5. However, we note that this
time estimate requires ignoring the first peaks in the data and comparing with slice plots
of the data. The core of the flux tube remains trapped, likely due to increased drag from
the larger cross section, so the mass flux is smaller than H7e17\_5. Because of the larger
size of the flux tube, the integrated mass flux peaks later than H7e17\_5 and is only
${\sim}20\%$ smaller. However, it is harder to distinguish the results from surface
oscillations, as seen in the top half of Fig.~\ref{fig:ejecta_big}. This
contamination also affects the additional integration surfaces (not shown), so the
velocity at emergence is significantly more uncertain. Nevertheless, we estimate it to be
${\sim}0.013c$, approximately a third of H7e17\_5 and H7e17\_0. The energetics for
H7e17\_5\_B are shown in the bottom plot of Fig.~\ref{fig:ejecta_big}. Qualitatively, the
energetics are similar to H7e17\_5, including the presence of a crossover around the time
of emergence.

\subsection{Effects of resolution}
\label{subsec:resolution}
To assess the effects of resolution, we include two additional tests for H7e17\_5 at
effective resolutions of ${\sim}79~\mathrm{m}$ (which we label ``LR'') and
${\sim}44~\mathrm{m}$ (or ``MR'') at the location of the flux tube, which correspond to
roughly $9$ and $16$ cells across the flux tube, respectively. The standard run at
${\sim}22~\mathrm{m}$ is denoted as ``HR'' for these comparisons. The LR run does not
emerge at all, as the flux tube expands horizontally too rapidly and stalls near the
surface of the star. As shown in Fig.~\ref{fig:ejecta}, the MR run features a slightly
delayed emergence time with a smaller mass flux compared to the HR run reported in
Table~\ref{tab:ics}. Similar to the LR case, this appears to be due to the tube spreading
more horizontally, which reduces how much of the flux tube fully emerges.

\begin{figure}
  \centering
  \includegraphics[width=\columnwidth]{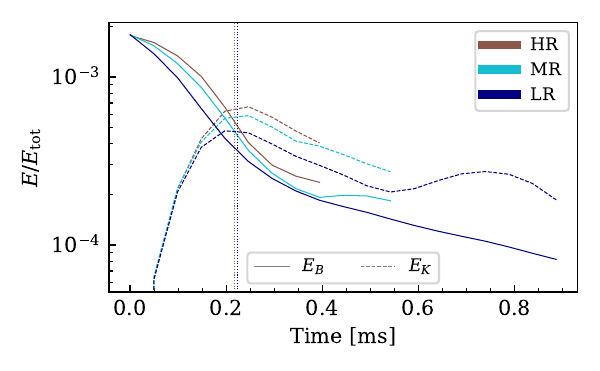}
  \caption{\label{fig:energetics_res} Like Fig.~\ref{fig:energetics}, but for the
  different resolutions of H7e17\_5. Dotted vertical lines mark the emergence time.}
\end{figure}

According to Fig.~\ref{fig:energetics_res}, the energetics are qualitatively similar
between the different resolutions, though the lower resolutions see $E_B$ decay faster and
less growth in $E_K$, consistent with additional diffusion in the flux tube. The crossover
point between the two exists in all three runs, including the LR run that does not emerge,
and seems to move earlier in time as the resolution decreases. As mentioned earlier,
however, the MR run emerges slightly later than HR, and LR does not emerge at all.
This suggests that the crossover time is not directly related to the emergence time, and
the agreement between the two in the HR run is coincidence.

Buoyancy should depend on resolving not only the flux tube but also the density and
pressure scale heights, $H_\rho \equiv -(d\rho/dr)^{-1}\rho$ and
$H_P \equiv -(dP/dr)^{-1}P$, respectively. Even if $H_\rho$ or $H_P$ are initially
resolved, both scale heights will decrease as we approach the surface and become
unresolved earlier and earlier as we decrease the resolution. We therefore also perform a
resolution study for H7e17\_5\_B to assess the effects of under-resolving $H_\rho$ and
$H_P$ more and more while the flux tube remains adequately resolved. In addition to LR and
MR configurations, we also add a very low resolution (VLR) configuration with
$\Delta r\approx 158~\mathrm{m}$ at the location of the flux tube. Therefore, the VLR, LR,
MR, and HR runs have ${\sim}18$, ${\sim}37$, ${\sim}67$, and ${\sim}134$ cells across the
flux tube. In all cases, part of the flux tube emerges successfully, though the core with
the highest magnetization remains trapped. The mass flux increases monotonically with
resolution except for the VLR case (see Fig.~\ref{fig:ejecta_big}), which appears to
push out noticeably more mass than any other resolution. However, as in H7e17\_5, the
integrated mass decreases rapidly across the other integration surfaces (not shown), even
for VLR, suggesting that it is still bound.

There is little consistency in the emergence time, and the effects of surface oscillations
make it difficult to determine exactly where emergence occurs, particularly for the LR
case. Though the flux tube is well resolved for the LR, MR, and HR cases (and arguably
adequately resolved for the VLR case), it is likely that we are not in a completely
convergent regime, even if we ignore the measured emergence times.

The energetics in Fig.~\ref{fig:ejecta_big} corroborate this picture; the behavior for
$E_B$ and $E_K$ are not entirely consistent with resolution, particularly for the VLR
case, which has a slightly larger initial $E_B$, and $E_K$ grows larger than both MR and
LR before decaying more rapidly (which is expected). However, the qualitative dynamics are
similar between all cases, and the predicted differences in emergence time, though
noticeable on the scale of our simulations ($t\sim1~\mathrm{ms}$ or less), would be small
over the course of a full BNS merger.

\section{Discussion}
\label{sec:discussion}
For our magnetic field configuration, we observe that only extremely large fields are
capable of emerging. Because our field is a single twisted loop surrounding the entirety
of the star, however, there is a strong magnetic tension force which, using
Eq.~\ref{eq:gr_inequality}, initially exceeded $50\%$ of the gravitational force on the
flux tube in some of our tests. Alternate field configurations may be less susceptible to
this tension force, but numerical simulations suggest that post-merger fields, independent
of their initial (realistic) conditions, are dominated by a large-scale toroidal component
\cite{Aguilera-Miret:2021fre,Gutierrez:2025gkx}, which limits the possibilities for
alternate field structures.

Even if this were not the case, this likely is only important for fields of
${\sim}10^{17}~\G$; we observe that the evolution of the flux tube at weaker field
strengths around $10^{16}~\G$ are dominated by ordinary hydrodynamic buoyancy. Therefore,
it seems unlikely that altering the field configuration is enough to help these features
to emerge. As stated in Sec.~\ref{subsec:id}, a rotating model is also unlikely to improve
emergence, as rotation tends to inhibit convective processes, not enhance them
\cite{1978RSPTA.289..459A,Spruit:1999cc,Gao:2025nfj}.

This result may initially seem at odds with works suggesting that strong magnetically
driven outflows may develop from a BNS remnant, such as \citet{Most:2023sft} and
\citet{Combi:2023yav}. This is not necessarily the case; the demonstrated flare-like
features and outflows in \citet{Most:2023sft} originate from strong magnetic fields
exceeding $10^{17}~\G$. Similarly, the post-merger jet in \citet{Combi:2023yav} develops
in a remnant with an average toroidal field strength nearing and isolated pockets
exceeding $10^{17}~\G$. However, we again stress that such field strengths are atypically
large compared to observation of magnetars and what is expected in realistic merger
remnants. Even under these circumstances, such fields are not a guarantee that
magnetically-driven outflows develop from the remnant. \citet{Musolino:2024sju}, for
example, suggest that, even with an extremely large magnetic field, these outflows may
instead originate from the disk.

Similar to alternate configurations, buoyant instabilities like the Parker instability are
likely only important for a small range of field strengths in excess of $10^{17}~\G$.
Despite a strong longitudinal perturbation, the Parker instability does not appear to
develop in H1e17\_5; rather, the flux tube oscillates. This agrees with results from
Refs.~\cite{Musolino:2024sju,Jiang:2025ijp}, for instance, which suggest that the
post-merger disk can be unstable to the Parker instability, but the remnant itself should
be stable, even for a very large magnetic field. Larger fields like H7e17\_5 may be
Parker-unstable, but the emergence time scale is so short that the instability cannot
develop quickly enough to be dynamically important (as seen when compared with H7e17\_0).
There likely exists some middle ground between these two states where the Parker
instability develops quickly enough to accelerate the emergence time scale, but the
required field strengths may not be achievable in a realistic merger setting.

As we consider cases which do emerge, it is interesting to note that whether or not
emergence occurs seems to depend only mildly on resolution so long as the buoyant feature
is adequately resolved. Though this is a necessary condition for emergence, one would not
necessarily expect it to be sufficient; under-resolving $H_\rho$ could lead to an
incorrect density difference between the ambient medium and the interior of the buoyant
feature, therefore leading to an incorrect emergence time scale. Similarly,
under-resolving $H_P$ could cause the buoyant feature to expand too much or too little as
it rises, thus indirectly affecting the density difference and limiting buoyancy. In our
runs, $H_\rho \approx 2.43~\km$ and $H_P \approx 827~\km$ at the shear layer, but both
these quantities will rapidly decrease as the flux tube approaches the surface. Therefore,
$H_\rho$ will only be moderately resolved at the beginning for the VLR run, and $H_P$ will
only be resolved by the HR and possibly MR runs to start. Though we see some differences
in the emergence time (which may not be entirely accurate for H7e17\_5\_B), they are small
enough that they would be insignificant in the context of a full merger simulation. This
suggests that the timescales shown in global BNS runs with
$\Delta x \gtrsim 180~\mathrm{m}$ are likely reasonable despite the limited resolution.

Secondary characteristics such as the amount of mass pushed out during emergence are more
strongly affected by resolution. Though our evidence is suggestive that these outflows
remain bound, it is not definitive. Furthermore, such outflows in a realistic merger would
also carry some angular momentum and interact with the magnetic fields already existing
outside the remnant. A scenario where some of this material becomes unbound is not
inconceivable, though how strong of an effect this is would be determined best by
high-resolution global simulations.

\section{Conclusion}
\label{sec:conclusion}
In this paper, we have investigated the dynamics of magnetically buoyant flux tubes inside
a neutron star remnant. We began by reviewing the concepts of magnetic buoyancy in a
Newtonian context, though we have also derived a semi-relativistic buoyancy condition for
a TOV spacetime as well. For a neutron star, a strong magnetic tension force appears due
to the toroidal structure of the magnetic field. This tension force inhibits buoyancy and
leads to an indirect condition on the equation of state for emergence to be possible, even
in the absence of stable stratification. Relativistic enhancements to gravity often relax
this constraint sufficiently for our remnant profile that flux tubes with realistic
post-merger fields are all at least marginally buoyant. Nevertheless, we note that this
tension force means that intuition from similar emergence problems in solar physics is
limited.

While our model only uses a simple, idealized field configuration, it captures the
essential features of a more realistic post-merger field. We additionally impose thermal
equilibrium rather than adiabaticity on our initial data, which increases the buoyancy of
the flux tube. Therefore, our results overestimate how buoyant a realistic post-merger
field would be.

Nevertheless, we also observe that these flux tubes can only overcome the stably
stratified profile of the remnant if the magnetic field is significantly more than
$10^{17}~\G$, which is often considered unphysically large. In agreement with
\citet{Musolino:2024sju}, this suggests that the strong magnetically-driven outflows
needed for BNS remnants to power relativistic jets may originate in the disk rather than
the remnant.

Though our work suggests emergence itself is somewhat independent of resolution so long as
the large-scale magnetic field is well resolved, secondary features like the mass flux may
be more sensitive to resolution. Therefore, a natural extension of this work is to explore
emergence with high-resolution global simulations where potential outflows can interact
with the remnant disk and flows. However, it is not immediately clear if such work would
be useful; if a BNS merger with a realistic initial field strength cannot
self-consistently develop a field well above $10^{17}~\G$ or demonstrate emergence at
lower field strengths, high-resolution simulations may be of little practical worth.

A possible study aimed toward probing emergence at lower field strengths could consider
the effects of the EOS; DD2 is a so-called ``stiff'' EOS, where a small change in density
produces a large shift in pressure, thus supporting heavier remnants with larger radii.
Due to the higher compactness, intuitively one would expect flux emergence in a long-lived
remnant with a softer EOS to be less likely because of increased magnetic tension.
However, enforcing pressure equilibrium would also lead to a larger relative difference in
density and enhance buoyancy, so it is not immediately obvious that a softer EOS might not
be more susceptible to buoyant effects.

\begin{acknowledgments}
The authors thank Elias Most, Patrick Cheong, and Eduardo M. Guti\'errez for helpful
discussions.
JF and DR acknowledge support from the U.S. Department of Energy, Office of Science, Division of
Nuclear Physics under Award Numbers DE-SC0021177.
DR also acknowledges funding from the National Science Foundation under
Grants No.~PHY-2020275, AST-2108467, PHY-2407681, the Sloan Foundation,
and from the U.S. Department of Energy, Office of Science, Division of
Nuclear Physics under Award Number DE-SC0024388.
PH acknowledges funding from the National Science Foundation under Grant No. PHY-2116686.
This work was completed in part at the TACC Open Hackathon, part of the Open Hackathons
program. The authors would like to acknowledge OpenACC-Standard.org for their support.
We are particularly grateful to Matthew Cawood, Victor Eijkhout, and Forrest Glines for
their assistance optimizing \AthenaK.
This research used resources of the National Energy Research Scientific Computing Center,
a DOE Office of Science User Facility supported by the Office of Science of the U.S.
Department of Energy under Contract No. DE-AC0205CH11231.
\end{acknowledgments}

\appendix
\onecolumngrid
\section{Equation of State Slices}
\label{app:fits}
\begin{figure}
  \centering
  \includegraphics[width=0.45\columnwidth]{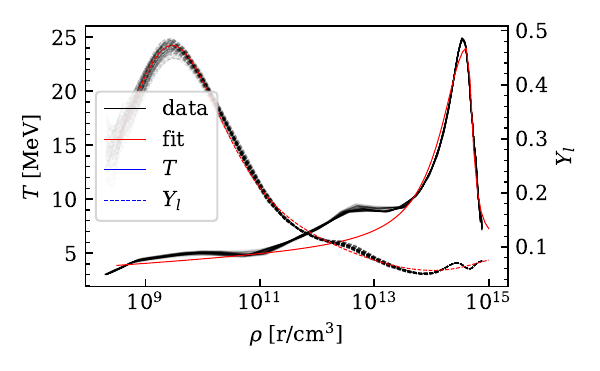}
  \includegraphics[width=0.45\columnwidth]{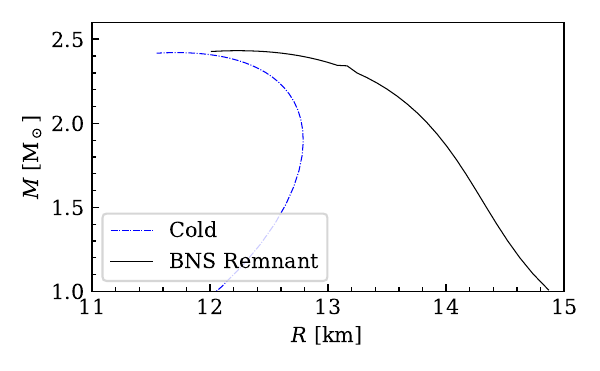}
  \caption{\label{fig:mr_curve} (Left) Fits of $T$ and $Y_l$ as functions of $\rho$ to the
  remnant from Ref.~\cite{Radice:2023zlw}. Black lines indicate profiles along the z-axis
  from the last $4.7$~\ms of the remnant, the red line marks the fit to this line. (Right)
  Mass-radius curves for the hot DD2 profile used in this work and the standard cold
  beta-equilibrium slice of DD2.}
\end{figure}

For the hot remnant profile, we perform nonlinear fits to the temperature $T$ and lepton
fraction $Y_l$ using the models
\begin{subequations}
\label{eq:fits}
\begin{align}
\label{eq:temp_model}
T(x) &= A + Bx +
  C D \left[D + \left(\frac{x - x_0}{1 - F\sign\left(x-x_0\right)}\right)^2\right]^{-1},\\
Y_l(x) &= \sum_{i=0}^7 c_i x^i,
\end{align}
\end{subequations}
where $x = \log{\rho}$. 
We use 100 profiles of the $z$-axis
equally spaced in time from the last $4.7$~ms of the remnant studied in
Ref.~\cite{Radice:2023zlw} (${\sim}100~\mathrm{ms}$ in duration).
For fits of $Y_l$, we compute $Y_l=Y_e + Y_{\nu_e} - Y_{\bar{\nu}_e}$, where $Y_e$,
$Y_{\nu_e}$, and $Y_{\bar{\nu}_e}$ are the electron, electron neutrino, and electron
antineutrino fractions, respectively. These fits are shown in the left half of
Fig.~\ref{fig:mr_curve}.

This profile is extrapolated to high density based on these fits. This is necessary
because the source profile is rotating, so the TOV case must be more compact to achieve a
high-mass remnant (which is essential so that the shear layer is close enough to the
surface to achieve buoyancy).


The mass-radius curve is shown in the right half of Fig.~\ref{fig:mr_curve}. For
reference, we also plot the standard cold beta-equilibrated slice of DD2.


\section{Relativistic Magnetic Buoyancy in the TOV Spacetime}
\label{app:gr_buoyancy}
\textit{Note: for this section, we consider geometrized units with $G=c=1$ and adopt
the Heaviside-Lorentz convention for the magnetic field; that is,
$B^i_\mathrm{HL} = B^i_\mathrm{CGS}/\sqrt{4\pi}$. We also use the Einstein summation
convention for paired indices, e.g., $a^\mu b_\mu \equiv \sum_\mu a^\mu b_\mu$.}

Consider the relativistic generalization of Eq.~\ref{eq:newt_v_lagrange}
\cite{Lichnerowicz:1994uzc}:
\begin{equation}
  \label{eq:gr_v_lagrange}
  \left(e + P + b^2\right)\frac{du^\alpha}{d\tau} +
    \mathcal{P}^\alpha_\nu g^{\mu\nu}\partial_\mu\left(P + \frac{b^2}{2}\right) -
    \mathcal{P}^\alpha_\nu \nabla_\mu\left(b^\mu b^\nu\right) =
    -\left(e + P + b^2\right)\Gamma^\alpha_{\mu\nu} u^\mu u^\nu,
\end{equation}
where $e$ is the total energy density, $P$ is the pressure, $b^\mu$ is the magnetic field
as measured in the comoving frame, $u^\alpha$ is the four-velocity, $\tau$ is the proper
time, and $\mathcal{P}^{\mu\nu} = g^{\mu\nu} + u^\mu u^\nu$ is a projection operator.
Assume there exists a state in hydrostatic equilibrium with energy density $e_0$, pressure
$P_0$, and no magnetic field. Then it follows that
\begin{equation}
  g^{\alpha i}\partial_i P_0 = -\left(e_0 + P_0\right) \Gamma^\alpha_{tt}u^t u^t.
\end{equation}
Now consider a state with $P^\ast = P_0 - \Delta P$, $e^\ast = e_0 - \Delta e$, and a
magnetic field $b^\mu$. We again enforce $P_0 = P^\ast + b^2/2$, which implies
$\Delta P = b^2/2$. If we consider the initial moment when the fluid is still at rest
($u^i = 0$, but derivatives of $u^\mu$ do not vanish), then Eq.~\ref{eq:gr_v_lagrange}
becomes
\begin{equation}
  \label{eq:gr_buoyant}
  \left(e^\ast + P^\ast + b^2\right)\frac{du^\alpha}{d\tau} -
    \nabla_\mu\left(b^\mu b^\alpha\right) - u^\alpha u_\nu b^\mu \nabla_\mu b^\nu =
    \left(\Delta e - \frac{b^2}{2}\right)\Gamma^\alpha_{tt}u^t u^t.
\end{equation}
We now focus exclusively on $u^r$, which eliminates the third term on the left-hand side.
Expanding the covariant derivative leads to
\begin{equation}
  \left(e^\ast + P^\ast + b^2\right)\frac{du^r}{d\tau} -
    \partial_\mu\left(b^\mu b^r\right) - \Gamma^\mu_{\mu\nu}b^\nu b^r -
    \Gamma^r_{\mu\nu}b^\mu b^\nu =
    \left(\Delta e - \frac{b^2}{2}\right)\Gamma^r_{tt}u^t u^t.
\end{equation}
As in the Newtonian case, we assume only the azimuthal field, $b^\phi$, is nonzero (note
that the $b^t$ term vanishes if $u_i=0$). This causes the
$\partial_\mu\left(b^\mu b^r\right)$ term to vanish, leading to 
\begin{equation}
  \label{eq:gr_tube_buoyancy}
  \left(e^\ast + P^\ast + b^2\right)\frac{du^r}{d\tau}-\Gamma^r_{\phi\phi}b^\phi b^\phi =
    \left(\Delta e - \frac{b^2}{2}\right)\Gamma^r_{tt}u^tu^t.
\end{equation}
In the TOV spacetime, $\Gamma^r_{\phi \phi} = -\left(1 - 2m(r)/r\right)r\sin^2\theta$
and $\Gamma^r_{tt} = \left(1- 2m(r)/r\right) \alpha \partial_r \alpha$, where $\alpha$ is
the lapse function. Since $b^\phi$ is the only nonzero magnetic field component,
$b^2 = \left(b^\phi\right)^2r^2\sin^2\theta$. Lastly, $u^0 u^0 g_{00} = -1$ implies that
$u^0 = 1/\sqrt{-g_{00}}$, so $u^0 = 1/\alpha$. Therefore,
\begin{equation}
  \frac{b^2}{r} < \left(\Delta e - \frac{b^2}{2}\right)\frac{\partial_r \alpha}{\alpha}.
\end{equation}
From the TOV equations, 
\begin{equation}
  \frac{d\alpha}{dr} = \frac{m(r) + 4\pi r^3 P_0}{r^2\left(1 - 2m(r)/r\right)}\alpha
\end{equation}
so
\begin{equation}
  \label{eq:gr_inequality}
  b^2 < \left(\Delta e - \frac{b^2}{2}\right)\frac{m(r) + 4\pi r^3 P_0}{r - 2m(r)}
\end{equation}

We will now show that this reduces to the Newtonian limit. First, recall that
$b^2/2 = \Delta P$ and $e = \rho + \rho\epsilon$. Then
$\Delta e - b^2/2 = \Delta\rho + \Delta(\rho\epsilon) - \Delta P$. Adding back in the
missing factors of $c^2$, the right-hand side of Eq.~\ref{eq:gr_inequality} becomes
\begin{equation}
  \left(\Delta\rho + \frac{1}{c^2}\Delta(\rho\epsilon) - \frac{\Delta P}{c^2}\right)
    \frac{m(r)}{r}\frac{\left(1 + 4\pi r^3 \frac{P_0}{m(r) c^2}\right)}
    {\left(1 - \frac{2m(r)}{rc^2}\right)}
\end{equation}
In the non-relativistic limit, $c\rightarrow\infty$ and $b^i\rightarrow B^i$, leaving us
with
\begin{equation}
  B^2 < \Delta \rho\frac{m(r)}{r}.
\end{equation}
Returning to CGS units, this yields
\begin{equation}
  B^2 < 4\pi \Delta\rho\frac{G m(r)}{r},
\end{equation}
which agrees with Eq.~\ref{eq:buoyant_condition}.

Eq.~\ref{eq:gr_inequality} is often a weaker constraint than its Newtonian
counterpart. A massive remnant may have $m(r)\sim2~\Msun$ near a shear layer located
at $r\sim7~G\Msun/c^2\approx10.3~\km$, suggesting that $(m(r) + 4\pi r^3 P_0)/(r - 2m(r))$
can be more than a factor of $2$ larger than $m(r)/r$. The precise behavior of the term
$\Delta e - \Delta P$ depends on the actual EOS, but a sufficient condition to ensure
that $\Delta e - \Delta P > \Delta\rho$ is that $d(\rho \epsilon)/dP > 1$. This is always
satisfied for an ideal gas respecting causality ($1 < \Gamma \leq 2$), though it may not
always hold for a realistic EOS. Nevertheless, we have found for our cases that the
product $(\Delta e - \Delta P)(1 + r\pi r^3 P_0/m(r))(1 - 2m(r)/r)^{-1} > 1$.



\bibliography{references}

\end{document}